\documentstyle[preprint,prabib,aps]{revtex}

\begin{document}
\draft
\date{\today}
\title{Warped, Anisotropic Wormhole/Soliton Configurations in Vacuum 5D Gravity }
\author{Sergiu I. Vacaru \thanks{
e-mails : sergiu$_{-}$vacaru@yahoo.com,\ sergiuvacaru@venus.nipne.ro}}
\address{Physics Department, CSU Fresno,\ Fresno, CA 93740-8031, USA, \\
and \\
Centro Multidisciplinar de Astrofisica - CENTRA, Departamento de Fisica,\\
Instituto Superior Tenico, Av. Rovisco Pais 1, Lisboa, 1049-001, Portugal}
\author{D. Singleton \thanks{
e-mail : dougs@csufresno.edu}}
\address{Physics Department, CSU Fresno,\ Fresno, CA 93740-8031, USA}
\maketitle

\begin{abstract}
In this paper we apply the anholonomic frames method developed in refs.
\cite{v,vst,vsbd,vtheorem} to construct and study anisotropic vacuum field
configurations in 5D gravity. Starting with an off--diagonal 5D metric,
parameterized in terms of several ansatz functions, we show that using
anholonomic frames greatly simplifies the resulting Einstein field
equations. These simplified equations contain an interesting freedom in that
one can chose one of the ansatz functions and then determine the remaining
ansatz functions in terms of this choice. As examples we take one of the
ansatz functions to be a solitonic solution of either the
Kadomtsev-Petviashvili equation or the sine-Gordon equation. There are
several interesting physical consequences of these solutions. First, a
certain subclass of the solutions discussed in this paper have an exponential
warp factor similar to that of the Randall-Sundrum model. However, the warp
factor depends on more than just the 5$^{th}$ coordinate. In addition the
warp factor arises from anisotropic vacuum solution rather than from any
explicit matter. Second, the solitonic character of these solutions might
allow them to be interpreted either as gravitational models for
particles ({\it i.e.} analogous to the ' t Hooft-Polyakov
monopole, but in the context of gravity), or as nonlinear, anisotropic
gravitational waves.
\end{abstract}

\vspace{1cm} PACS: 04.50.+h

\vspace{1cm}

\section{Introduction}

The study of wormhole and black hole solutions in general relativity
or its higher dimensional extensions usually starts with solutions
with a high degree of symmetry \cite{mor}, such as spherical symmetry.
These highly symmetric solutions generally have metrics which are
diagonal, and are thus easier to handle from a computational point
of view. However, it is of interest to consider situations with
less symmetry and metrics with off--diagonal components.
Salam, Strathee and Percacci \cite{sal} have shown that including
off--diagonal components in higher dimensional metrics is equivalent to
including gauge fields. A specific example of this is given in refs.
\cite{chodos,dzhsin} where 5D metrics with off--diagonal terms were
considered, leading to solutions which were similar
to spherically symmetric 4D wormhole or flux tube metrics
but with ``electric'' and/or ``magnetic'' fields running
along the throat of the wormhole. The ``electromagnetic'' fields arose
as a consequence of the off--diagonal elements of the 5D metric.

Most realistic, physical situations do not have the high degree
of symmetry that is found in the well known, exact solutions of
general relativity. For example, it would be somewhat surprising
if a pure Schwarzschild black hole existed ({\it i.e.} a black
hole without angular momentum or charge). In order to study
anisotropic configurations we use the method
of anholonomic frames developed in \cite{v,vst,vtheorem}
with associated nonlinear connections. The basic idea of this
method is that the anholonomic frames can diagonalize metrics,
which would have off--diagonal components in a holonomic
coordinate frame. This substantially
simplifies tensor computations but ``elongates'' the rule of partial
differentiation and results in an anholonomic dynamics of particle/field
interactions. Previously the anholonomic frames method was used
\cite{vsbd} to construct {\it anisotropic} wormhole and flux tube
solutions, which reduced to the solutions of \cite{chodos,dzhsin,ds} in the
isotropic limit. These anisotropic solutions exhibited a variation or
running of the ``electromagnetic'' charges as a result of the
angular anisotropies and/or through variations of the extra spatial
dimension. The solutions of ref. \cite{vsbd} were anisotropic
deformations from spherically symmetric solutions.
In ref. \cite{vs}, we showed that the
anholonomic frames method could be used to construct anisotropic
wormhole and flux tube solutions to 5D Kaluza--Klein theory which
possessed a host of different symmetries (elliptic, cylindrical,
bipolar, toroidal).

In this paper we extend the results of \cite{vsbd,vs} to show that
these locally anisotropic wormhole and flux tube solutions
can be modified to include a class of metrics with a warped geometry
with respect to the 5$^{th}$ coordinate. In our approach the warp factor
is not generated by some brane configuration with
a special energy--momentum tensor as in the
Randall--Sundrum (RS) scenario \cite{rs}, but is induced by a specific
anholonomic frame relation and nonlinear polarization of the vacuum 5D
gravitational fields in the bulk. This supports the idea that off--diagonal
metrics and anisotropies in extra dimensional vacuum gravity can model low
dimensional interactions for both gravity and matter
(more details can be found in \cite{vgg} ). We will
discuss the physical consequences of these solutions, in particular
the gravitational running of the electromagnetic constants that
occurs, and the anisotropic and solitonic nature of
the electromagnetic and gravitational fields.

The paper is organized as follows: In section II we introduce an
off--diagonal ansatz for 5D vacuum gravitational fields and outline the
anholonomic frame method for this ansatz by computing the nontrivial
components of the Ricci tensor and constructing the general 5D vacuum
solutions for Einstein's equations. The coupled non-linear equations for the
ansatz functions have a freedom that allows us to specify a particular form
for one of the functions and then determine the other ansatz functions for
this choice. We use this freedom to construct solutions that have one of the
ansatz functions associated with 3D solitons. In section III we combine
these solitonic solutions with certain 5D, ``electric/magnetic'' charged
wormhole and flux tube solutions. The result
is a {\it vacuum} solution with an
anisotropic RS type warp factor. In section IV we show that it is possible
to introduce an anisotropic deformation of these solutions so that the
``electric'' and ``magnetic'' charges of the solutions develop an
anisotropic dependence. In section V we show that in addition to the
spherical 3D hypersurface backgrounds that we used as a starting point for
building the solitonically deformed wormhole and flux tube solutions, it is
also possible to consider other 3D background geometries (ellipsoidal,
toroidal, bipolar or cylindrical). Conclusions and physical consequences of
the various solutions are given in section VI.

\section{A Class of 5D Vacuum Gravitational Fields}

In this section we outline the method of anholonomic frames with application
to an ansatz for 5D vacuum gravitational metrics. These metrics have a
mixture of holonomic and anholonomic variables, and are most naturally dealt
with using anholonomic frames (the general theorems and corollaries are
formulated and proven in refs. \cite{vtheorem,vgg}). We integrate the vacuum
Einstein equations in explicit form and analyze the physical and
mathematical properties of the 5D locally anisotropic vacuum solutions which
can be reparametrized to yield warped and solitonic wormhole -- flux tube
configurations.

\subsection{Off--diagonal metric ansatz and anholonomic Einstein equations}

Let us consider a 5D pseudo--Riemannian spacetime of signature $(+,-,-,-,-)$
and denote the local coordinates $u^{\alpha
}=(x^{i},y^{a})=(x^{1}=t,x^{2}=r,x^{3}=\theta ,y^{4}=\chi ,y^{5}=\varphi )$.
$\chi $ is the 5$^{th}$ coordinate, $t$ is the time like coordinate and
$(r,\theta ,\varphi )$ are the usual spherical spatial coordinates. The Greek
indices are split into two subsets $x^{i}$ (holonomic coordinates) and $y^{a}
$ (anholonomic coordinates) which are labeled respectively by Latin indices
$i,j,k,...=1,2,3$ and $a,b,...=4,5$. The local coordinates will sometimes be
written in the compact form $u=(x,y)$.

We begin our approach by considering a 5D quadratic line element
\begin{equation}
dS^{2}=\Omega ^{2}\left( x^{i},\chi \right) g_{\alpha \beta }\left(
x^{i},\chi \right) du^{\alpha }du^{\beta }  \label{cmetric}
\end{equation}
where $\Omega ^{2}\left( x^{i},\chi \right)$ is a conformal factor. The
metric coefficients $g_{\alpha \beta }$ are parametrized by an off--diagonal
matrix ansatz {\scriptsize
\begin{equation}
\left[
\begin{array}{ccccc}
1+(w_{1}^{\ 2}+\zeta _{1}^{\ 2})h_{4}+n_{1}^{\ 2}h_{5} & (w_{1}w_{2}+\zeta
_{1}\zeta _{2})h_{4}+n_{1}n_{2}h_{5} & (w_{1}w_{3}+\zeta _{1}\zeta
_{3})h_{4}+n_{1}n_{3}h_{5} & (w_{1}+\zeta _{1})h_{4} & n_{1}h_{5} \\
(w_{1}w_{2}+\zeta _{1}\zeta _{2})h_{4}+n_{1}n_{2}h_{5} & g_{2}+(w_{2}^{\
2}+\zeta _{2}^{\ 2})h_{4}+n_{2}^{\ 2}h_{5} & (w_{2}w_{3}++\zeta _{2}\zeta
_{3})h_{4}+n_{2}n_{3}h_{5} & (w_{2}+\zeta _{2})h_{4} & n_{2}h_{5} \\
(w_{1}w_{3}+\zeta _{1}\zeta _{3})h_{4}+n_{1}n_{3}h_{5} & (w_{2}w_{3}++\zeta
_{2}\zeta _{3})h_{4}+n_{2}n_{3}h_{5} & g_{3}+(w_{3}^{\ 2}+\zeta _{3}^{\
2})h_{4}+n_{3}^{\ 2}h_{5} & (w_{3}+\zeta _{3})h_{4} & n_{3}h_{5} \\
(w_{1}+\zeta _{1})h_{4} & (w_{2}+\zeta _{2})h_{4} & (w_{3}+\zeta _{3})h_{4}
& h_{4} & 0 \\
n_{1}h_{5} & n_{2}h_{5} & n_{3}h_{5} & 0 & h_{5}
\end{array}
\right]  \label{ansatz0}
\end{equation}
} where the coefficients are smooth functions of the form:
\[
g_{2,3}=g_{2,3}(x^{2},x^{3}) \; \;, \; \; h_{4,5}=h_{4,5}(x^{i},\chi ) \; \;
, \; \; w_{i}=w_{i}(x^{i},\chi ) \; \; , \; \; n_{i}=n_{i}(x^{i},\chi ) \;
\; , \; \; \zeta _{i}=\zeta_{i}\left( x^{i},\chi \right) .
\]
The quadratic line element (\ref{cmetric}) with metric coefficients (\ref
{ansatz0}) can be diagonalized,
\begin{equation}
\delta S^{2}=\Omega ^{2}[(dt)^{2}+g_{2}(dr)^{2}+g_{3}(d\theta )^{2}+h_{4}
(\widehat{\delta }\chi )^{2}+h_{5}(\delta \varphi )^{2}],  \label{dmetric}
\end{equation}
with respect to the anholonomic co--frame $\left( dx^{i},\widehat{\delta }
\chi,\delta \varphi \right) ,$ where
\begin{equation}
\widehat{\delta } \chi=d \chi +\left( w_{i}+\zeta _{i}\right) dx^{i} \;
\mbox{and } \; \delta \varphi =d \varphi +n_{i}dx^{i}  \label{ddif1}
\end{equation}
which is dual to the frame $\left( \widehat{\delta }_{i},
\partial_{\chi},\partial _{\varphi}\right) ,$ where
\begin{equation}
\widehat{\delta }_{i}=\partial _{i}+\left( w_{i}+\zeta _{i}\right) \partial
_{\chi}+n_{i}\partial _{\varphi}.  \label{dder1}
\end{equation}
The various partial derivatives are abbreviated as $\dot{a} =\partial
/\partial t$ ,$a^{\bullet }=\partial /\partial r$, $a^{\prime} =\partial
/\partial \theta$ , $a^{\ast }=\partial /\partial \chi.$ The first order
anisotropy of metric (\ref{cmetric}) with ansatz (\ref{ansatz0}) comes
from the coefficients $w_{i}$ and $n_{i}$, while any second order anisotropy
comes from having non-trivial coefficients $\zeta _{i}$ (details can be
found in \cite{vtheorem} and \cite{vst,vgg}). For the trivial conformal
factor $\Omega =1$ in (\ref{cmetric}) and (\ref{dmetric}) there is only a
first order anisotropy parametrized by ansatz (\ref{ansatz0}) with $\zeta
_{i}=0.$

The nontrivial components of the 5D vacuum Einstein equations, $R_{\alpha
}^{\beta }=0,$ for the metric (\ref{dmetric}) given with respect to
anholonomic frames (\ref{ddif1}) and (\ref{dder1}) are:
\begin{eqnarray}
R_{2}^{2}=R_{3}^{3}=-\frac{1}{2g_{2}g_{3}}[g_{3}^{\bullet \bullet }-\frac{
g_{2}^{\bullet }g_{3}^{\bullet }}{2g_{2}}-\frac{(g_{3}^{\bullet })^{2}}{
2g_{3}}+g_{2}^{^{\prime \prime }}-\frac{g_{2}^{^{\prime }}g_{3}^{^{\prime }}
}{2g_{3}}-\frac{(g_{2}^{^{\prime }})^{2}}{2g_{2}}] &=&0,  \label{ricci1a} \\
R_{4}^{4}=R_{5}^{5}=-\frac{\beta }{2h_{4}h_{5}} &=&0,  \label{ricci2a} \\
R_{4i}=-w_{i}\frac{\beta }{2h_{5}}-\frac{\alpha _{i}}{2h_{5}} &=&0,
\label{ricci3a} \\
R_{5i}=-\frac{h_{5}}{2h_{4}}\left[ n_{i}^{\ast \ast }+\gamma n_{i}^{\ast } 
\right] &=&0,  \label{ricci4a} \\
\partial _{i}\ln |\Omega |-\left( w_{i}+\zeta _{i}\right) (\ln |\Omega
|)^{\ast } &=&0,  \label{confeq}
\end{eqnarray}
where
\begin{eqnarray}
\alpha _{i} &=&\partial _{i}{h_{5}^{\ast }}-h_{5}^{\ast }\partial _{i}\ln
\sqrt{|h_{4}h_{5}|},  \label{abc} \\
\beta &=&h_{5}^{\ast \ast }-h_{5}^{\ast }[\ln \sqrt{|h_{4}h_{5}|}]^{\ast },
\nonumber \\
\gamma &=&\frac{3h_{5}^{\ast }}{2h_{5}}-
\frac{h_{4}^{\ast }}{h_{4}}.  \nonumber
\end{eqnarray}
The anholonomic frames method has reduced the Einstein field equations for
the complex, off-diagonal, 5D metric of (\ref{ansatz0}) to a relatively
``simple'' system of coupled, nonlinear (\ref{ricci1a}) -- (\ref{confeq}).
In the following sections we will show that this simplification allows us to
construct exact solutions to these 5D vacuum equations by embedding well
known solitonic configurations into the system via the ansatz function
$h_4$, $h_5$ or $n_i$.

\subsection{Solutions of anholonomic vacuum Einstein equations}

The system of equations (\ref{ricci1a})--(\ref{confeq}) has to some extent
decoupled the equations for some of the ansatz functions through the use of
anholonomic frames. Thus one has some freedom to chose one or more of the
ansatz functions to take a particular, interesting form, and then solve for
the remaining functions. For example one could chose $g_2 (r, \theta)$ and
then use equation (\ref{ricci1a}) to solve for $g_3 (r, \theta)$

\begin{itemize}
\item  One solution of equation (\ref{ricci1a}) is
\[
g_{1}(r,\theta )=g_{2}(r,\theta )=g(r,\theta )=g_{[0]} + g_{[1]}\exp[%
a_{2}r+a_{3}\theta ],
\]
were $g_{[0]},a_{2}$ and $a_{3}$ are some constants. This can be generalized
by considering a coordinate transformation from $r,\theta \rightarrow
\widetilde{x}^{2,3}\left( r,\theta \right) $ so that the 2D line element
becomes conformally flat
\[
g_{2}(r,\theta )(dr)^{2}+g_{3}(r,\theta )(d\theta )^{2}\rightarrow
g(r,\theta )\left[ (d\widetilde{x}^{2})^{2}+\epsilon (d\widetilde{x}
^{3})^{2} \right]
\]
this is always possible for 2D metrics. The explicit form of $\widetilde{x}
^{2,3}$ depends on the signature $\epsilon =\pm 1.$ With this transformation
we write
\[
g_{1,2}(r,\theta ) =g_{1,2[0]}+g_{1,2[1]}\exp [a_{2}\widetilde{x}^{2}\left(
r,\theta \right) +a_{3}\widetilde{x}^{3}\left( r,\theta \right) ],
\]
for constants $g_{1,2[0]}$ and $g_{1,2[1]}.$ The simplest case for solutions
to equation (\ref{ricci1a}) is when the ansatz functions only depend on a
single variable such as $g_{2}(r)$ and $g_{3}(\theta )$ so that the
derivatives automatically vanish.

\item  Equation (\ref{ricci2a}) relates two functions $h_{4}\left(
x^{i},\chi \right) $ and $h_{5}\left( x^{i},\chi \right) $. There are two
possibilities. First if $h_{4}\left( x^{i},\chi \right)$ is given then
$h_{5}\left( x^{i},\chi \right)$ can be computed as

\begin{eqnarray}
\sqrt{|h_{5}|} &=&h_{5[0]}\left( x^{i}\right) +h_{5[1]}\left( x^{i}\right)
\int \sqrt{|h_{4}\left( x^{i},\chi \right) |}d\chi , \qquad
\mbox{for} \; \; h_{4}^{\ast}\left( x^{i},\chi \right) \neq 0;  \nonumber \\
&=&h_{5[0]}\left( x^{i}\right) +h_{5[1]}\left( x^{i}\right) \chi , \qquad
\mbox{for} \;\; h_{4}^{\ast }\left( x^{i},\chi \right) =0,  \label{p2}
\end{eqnarray}
where $h_{5[0,1]}\left( x^{i}\right) $ are functions determined by boundary
conditions. Second, if $h_{5}\left( x^{i},\chi \right)$ is given and
$h_{5}^{\ast }\neq 0$ then $h_{4}\left( x^{i},\chi \right)$ can be
computed as

\begin{equation}
\sqrt{|h_{4}|}=h_{[0]}\left( x^{i}\right) (\sqrt{|h_{5}\left( x^{i},\chi
\right) |})^{\ast },  \label{p1}
\end{equation}
with $h_{[0]}\left( x^{i}\right) $ given by boundary conditions.

\item  The exact solutions of (\ref{ricci3a}) are
\begin{equation}
w_{k}= \frac{\partial _{k}\ln
\left[\frac{\sqrt{|h_{4}h_{5}|}}{|h_{5}^{\ast }|}\right]}{\partial _{\chi}
\ln \left[ \frac{\sqrt{|h_{4}h_{5}|}}{|h_{5}^{\ast }|} \right]},
\label{w}
\end{equation}
for $\partial _{\chi }=\partial /\partial \chi $ and $h_{5}^{\ast }\neq 0.$
If $h_{5}^{\ast }=0,$ or even $h_{5}^{\ast }\neq 0$ but $\beta =0,$ the
coefficients $w_{k}$ could be arbitrary functions of $\left( x^{i},\chi
\right) .$ For vacuum Einstein equations this is a degenerate case which
imposes the compatibility conditions $\beta =\alpha _{i}=0,$ which are
satisfied if the $h_{4}$ and $h_{5}$ are related via formula (\ref{p1}) with
$h_{[0]}\left( x^{i}\right) =const.$ For simplicity, in this paper we will
consider only vacuum configurations with $w_{k}=0$.

\item  The exact solution of (\ref{ricci4a}) is
\begin{eqnarray}
n_{k} &=&n_{k[0]}\left( x^{i}\right) +n_{k[1]}\left( x^{i}\right) \int
\left[\frac{h_{4}}{(|h_{5}|)^{3/2}} \right]
d\chi , \qquad \mbox{for} \; \; h_{5}^{\ast }\neq 0;  \nonumber \\
&=&n_{k[0]}\left( x^{i}\right) +n_{k[1]}\left( x^{i}\right) \int h_{4}d\chi
, \qquad \mbox{for} \; \; h_{5}^{\ast }=0;  \label{n} \\
&=&n_{k[0]}\left( x^{i}\right) +n_{k[1]}\left( x^{i}\right) \int
\left[ \frac{1}{(\sqrt{|h_{5}|})^{3}} \right]d\chi \,
\qquad \mbox{for} \; \; h_{4}^{\ast }=0,
\nonumber
\end{eqnarray}
where the functions $n_{k[0,1]}\left( x^{i}\right) $ are determined from
boundary conditions.

\item  If $\partial_{i}\Omega =0$ and $\Omega ^{\ast }=0$ the exact solution
of (\ref{confeq}) is given by an arbitrary function $\zeta _{i}=\zeta
_{i}\left( x^{i},\chi \right)$. If $\Omega ^{\ast } \neq 0$ but $\partial
_{i}\Omega = 0$ we take $\zeta _{i}=0$. Finally if both $\Omega ^{\ast }
\neq 0$ and $\partial _{i}\Omega \neq 0$, $\zeta _i$ takes the form
\begin{equation}  \label{confsol}
\zeta _i =(\Omega ^{\ast })^{-1}\partial _{i}\Omega
\end{equation}
$w_i$ does not show up in this solution since in this paper we are
considering only the, $w_i =0$ case.
\end{itemize}

Because the metric coefficients $h_{4}$ and $h_{5}$ are solutions to
equation (\ref{ricci2a}) one has the freedom to define two
new solutions as $\widehat{h}_{4} =\eta _{4}h_{4}$,
$\widehat{h}_{5}=\eta _{5}h_{5}]$ where $\eta _{4,5}$ could depend
on $t, r, \theta$ or $\chi$. We call the functions $\eta_{4,5}
=\eta _{4,5}(t,r,\theta ,\chi )$ gravitational polarizations since
they modify the behavior of the metric coefficients $h_{4}$ and $h_{5}$ in a
manner similar to how a material modifies the behavior of electric and
magnetic fields in media. The ``renormalization'' of $h_{4,5}$ into
$\widehat{h}_{4,5}$ results in corresponding ``renormalizations''
$n_{i}\rightarrow \widehat{n}_{i},$ $\zeta _{i}\rightarrow \widehat{\zeta }
_{i}$ and $\Omega \rightarrow \widehat{\Omega }$ which are to be computed
from equations (\ref{p2})-- (\ref{confsol}) with redefined coefficients (\ref
{abc}), $\gamma \rightarrow \widehat{\gamma },\beta \rightarrow \widehat{
\beta },\alpha _{i}\rightarrow \widehat{\alpha }_{i}$. We will use the
freedom to renormalize $h_{4,5}$ via $\eta _{4,5}$ to modify certain
electric and magnetically charged, 5D wormhole and flux tube solutions.

\subsection{3D solitonic configurations embedded in 5D gravity}

Vacuum gravitational 2D solitons in 4D Einstein gravity were originally
studied by Belinski and Zakharov \cite{belinski}. In refs. \cite{v} it was
shown that there are 3D solitons in 4D gravity and in anisotropic Taub-NUT
backgrounds. Here we will show that it is possible to embed 3D solitonic
configurations into the 5D gravitational system of the preceding section.

The simplest way to construct a solitonic, off--diagonal 5D vacuum metric is
to take one of the ansatz coefficients $h_{4}$, $h_{5}$ or $n_i$ as a
solitonic solution of some particular non-linear equation. Then carrying out
the integrations of equations (\ref{p2})--(\ref{confsol}) yields all the
remaining ansatz functions. In the next two subsections we analyze two
explicit examples of such solitonic solutions. First we take either $h_4$ or
$h_5$ as solitonic and then determine $n_i$ in terms of this choice. Second,
we take one of the $n_i$'s as solitonic and determine $h_4$ and $h_5$.

\subsubsection{Solitonic $h_{4}$ or $h_{5}$}

The coefficient $h_{5}(r,\theta ,\chi )$ can be required to be a solution of
the Kadomtsev--Petviashvili (KdP) equation \cite{kad} or the (2+1) sine-Gordon
(SG) equation. Methods of dealing with the KdP and other 2+1 dimensional
soliton equations can be found in refs. \cite{dryuma} and \cite{zakhsh}
. More detailed information on the SG equation can be found in \cite
{har,lieb,whith}. In the KdP case $h_{5}(r,\theta ,\chi )$ satisfies the
following equation
\begin{equation}
h_{5}^{\ast \ast }+\epsilon \left( \dot{h}_{5}+6h_{5}h_{5}^{\prime
}+h_{5}^{\prime \prime \prime }\right) ^{\prime }=0,
\qquad \epsilon =\pm 1,
\label{kdp}
\end{equation}
while in the SG case $h_{5}(r,\theta ,\chi )$ satisfies
\begin{equation}
h_{5}^{\ast \ast }+h_{5}^{^{\prime \prime }}-\ddot{h}_{5}=\sin (h_{5}).
\label{sineq}
\end{equation}
We will use the notation $h_{5}=h_{5}^{KP}$ or $h_{5}=h_{5}^{SG}$ depending
on if $h_5$ satisfies equation (\ref{kdp}), or (\ref{sineq}) respectively.

Haven chosen a solitonic value of $h_{5}=h_{5}^{KP,SG},$ we can compute
$h_{4}$ by applying equation (\ref{p1}),
\begin{equation}
h_{4}=h_{4}^{KP,SG}=h_{[0]}^{2}
[(\sqrt{h_{5}^{KP,SG}})^{^{\ast }}]^{2}.  \label{p1b}
\end{equation}
The $h_{4}$ determined in this way will not necessarily share the solitonic
character of $h_{5}$. The next step is to substitute the values
$h_{4}^{KP,SG}$ and $h_{5}^{KP,SG}$ into the equation (\ref{n}) and find the
respective values $n_{i}=n_{i}^{KP,SG}(r,\theta ,\chi ).$ This, up to some
explicit integrations and differentiations, gives an exact solution of the 5D
vacuum Einstein equations generated from a solitonic equation for $h_{5}.$
For simplicity we will set $g_{2,3}=1$ so that the holonomic 2D background
is trivial.

In this subsection we have generated solutions to the system of equations
(\ref{ricci1a})--(\ref{confeq}) by requiring that $h_{5}$ satisfy some
solitonic equation. It is also possible to carry out a similar construction
but with $h_{4}$ satisfying equation (\ref{kdp}) or equation (\ref{sineq})).

\subsubsection{Solitonic n$_{i}$}

Next we consider the case when $n_i$ is the solitonic solution of either the
KdP or the SG equation. Specially we will take $n_3$ as the solution of
either the KdP or SG equation, setting the other two components to zero
($n_{1,2}=0$).

For $n_{3}$ satisfying the KdP equations
\[
n_{3}^{\ast \ast }+\epsilon \left( n_{3}^{\bullet }+6n_{3}n_{3}^{\prime
}+n_{3}^{\prime \prime \prime }\right) ^{\prime }=0,
\qquad \epsilon =\pm 1,
\]
in which case we write $n_{3}=n_{3}^{KP}$, while for $n_{3}$
satisfying the SG equation
\[
n_{3}^{\ast \ast }+n_{3}^{^{\prime \prime }}-n_{3}^{\bullet \bullet }=\sin
(n_{5}),
\]
we write $n_{3}=n_{3}^{SG}.$ We now want to find $h_{4}$ and
$h_{5}$ for the given solitonic ansatz $n_{3}=n_{3}^{KP,SG}.$ If
$h_{5}^{\ast}\neq 0$ then $h_{4}$ can be expressed in terms of $h_{5}$
using formula (\ref{p1b}) with $h_{[0]}=const$. The coefficient $\gamma$
can be expressed in terms of $h_{5}$ from (\ref{abc})
\begin{equation}
\gamma =\left( \ln \left[ \frac{|h_{5}|^{3/2}}{(h_{5}^{\ast })^{2}}\right]
\right) ^{\ast }.  \label{gamma1}
\end{equation}
Inserting (\ref{gamma1}) into equation (\ref{ricci4a}) with the one
nontrivial value $n_{3}=n_{3}^{KP,SG}$ we obtain an explicit equation for
$h_{5}$
\[
h_{5}^{KP,SG}=\left[ n_{[0]}\left( x^{i}\right) \int
\sqrt{|n_{3}^{KP,SG}(r,\theta ,\chi )|}
d\chi +n_{[1]}\left( x^{i}\right) \right]
^{4},
\]
were $n_{[0]}\left( x^{i}\right)$ and $n_{[1]}\left( x^{i}\right) $ are
fixed from boundary conditions. Once the explicit form of $h_{5}$ is given
via the above expression one can go back and obtain explicit expression for
$h_{4}$ and $\gamma$ from equations (\ref{p1}) and (\ref{abc}).

The main conclusion of this subsection is that the ansatz (\ref{ansatz0})
when treated with anholonomic frames has the freedom 
to pick one of the ansatz functions ($h_4$ , $h_5$ , or $n_i$) to satisfy
some 3D solitonic equation. Then in terms of this choice all the other
ansatz functions can be generated up to carrying out some explicit
integrations and differentiations. In this way it is possible to build exact
solutions of the 5D vacuum Einstein equations with a solitonic character.

There are other variations of the above solitonic constructions that could
be tried: instead of taking $h_{4,5}$ to depend on $r,\theta$ and $\chi$ we
could carry out the construction with other dependences ({\it e.g.} $%
h_{4}\left( t,\theta ,\chi \right) , h_{5}\left( t,\theta ,\chi \right),$ or
$h_{4}\left( t,r,\chi \right) , h_{5}\left( t,r,\chi \right),$ or $%
h_{4}\left( t,\chi \right) , h_{5}\left( t,\chi \right)$).

\section{Wormhole and Flux Tube Solutions with a Warp Factor}

In this section we analyze a subclass of metrics (\ref{dmetric}) with a
specific type of conformal factor parametrized as
\begin{equation}
\Omega \left( t,r,\theta ,\chi \right) =\Omega _{0}\left( t,r,\theta \right)
e^{-k\left( t,r,\theta \right) |\chi |}  \label{conf1}
\end{equation}
The functions $k\left( t,r,\theta \right) $ and $\Omega _{0}\left(
t,r,\theta \right) $ can be defined from boundary conditions or some desired
limit. In this paper we will assume that $k\left( t,r,\theta \right) $ and
$\Omega _{0}\left( t,r,\theta \right) $ are of the form of the radion or
modulus field from brane and string cosmology \cite{radion}. In the
original RS solution \cite{rs} the warp factor was of the form given in
equation (\ref{conf1}) but with $k\left( t,r,\theta \right) =const$. This
led to unstable  solutions and less realistic cosmological scenarios.
Some of these problems could be
avoided by allowing the constant in the exponential warp factor to
become a variable. A consequence of this was the introduction of the radion
or modulus field. More details of this can be found in refs. \cite{radion}
where different forms of the radion field, $k\left( t,r,\theta
\right)$, were proposed to generate anisotropic warped compactifications
or inflationary scenarios, which are compatible with string gravity or
other quantum gravity models.

From equation (\ref{confsol}) the parametrization in equation (\ref{conf1})
results in a second order anisotropy
\[
\zeta _{i}=\left( \Omega ^{\ast }\right) ^{-1}\partial _{i}\Omega
=-k_{\varepsilon } ^{-1} \left( t,r,\theta \right) \partial _{i}\ln |\Omega
_{0}|+|\chi |\partial _{i}\ln |k_{\varepsilon }\left( t,r,\theta \right) |
\]
$k_{\varepsilon }=k,$ if $\chi >0$ and $k_{\varepsilon }=-k,$ if $\chi <0$.
These $\zeta _{i}$'s can then be used to calculate the anholonomically
``elongated'' partial derivatives in (\ref{ddif1}) and (\ref{dder1}).

For simplicity, in this paper we shall deal only with exponential conformal
factors of the form
\begin{equation}
\Omega _{0}\simeq 1,k\simeq k\left( t\right) ,  \label{expconf}
\end{equation}
with a linear dependence on the 5$^{th}$ coordinate $\chi $
\[
\zeta _{1}\left( t,\chi \right) =|\chi |\partial _{t}\ln |k\left( t\right)
|,\zeta _{2}=\zeta _{3}=0,
\]
Putting these requirements together leads to a metric of the form
\begin{eqnarray}
\delta S^{2} &=&e^{-2k\left( t\right) |\chi |}[(dt)^{2}+g_{2}(r,\theta
)(dr)^{2}+g_{3}(r,\theta )(d\theta )^{2}  \label{dmetric1} \\
&&+\eta _{4}(t,r,\theta ,\chi )h_{4}(r,\theta )(\widehat{\delta }\chi
)^{2}+\eta _{5}(t,r,\theta ,\chi )h_{5}(r,\theta )(\delta \varphi )^{2}].
\nonumber
\end{eqnarray}
where we have inserted the gravitational polarizations, $\eta _{4,5}$.
Such a metric with its radion field warp factor is studied in connection
with brane and string cosmological models. The new feature in the solutions
that we construct is that they are vacuum solutions which do not have any
explicit matter.

In the following subsections we will show how the isotropic
wormhole/flux tube solutions of ref. \cite{dzhsin,ds} can be deformed
into the above anisotropic form with a warp factor and solitonic
background.

\subsection{5D isotropic wormholes and anisotropic solitons}

We give a brief review of the locally isotropic wormhole and flux tube
solutions constructed in refs. \cite{dzhsin,ds} (DS-solutions). Then in
stages we modify these locally isotropic solutions to incorporate the
solitonic ansatz function and the radion like warp factor.

\subsubsection{The locally isotropic DS--solutions}

The following form for the ansatz functions in ansatz (\ref{dmetric1})
\begin{eqnarray}
g_{1} &=&1 , \qquad g_{2}=-1 , \qquad g_{3}=-a(r) , \qquad
k=k_{t}=k_{r}=k_{\theta }=0,  \nonumber \\
h_{4} &=&-a(r)\sin ^{2}\theta , \qquad h_{5}=-r_{0}^{2}e^{2\psi (r)} ,
\qquad \eta _{4}=\eta_{5}=1,  \label{solds} \\
n_{1} &=& \omega (r),\qquad n_{2}=0,\qquad n_{3}=n_{[0]}\cos
\theta  \nonumber
\end{eqnarray}
defines a trivial, locally isotropic solution of the vacuum Einstein
equations (\ref{ricci2a})--(\ref{ricci4a}) which satisfies the conditions
$h_{4,5}^{\ast }=0.$ The analytic or numerical determined forms for
$a(r), \omega (r)$ and $\psi (r)$ are given in refs. \cite{dzhsin,ds}.
Inserting the functions from equation (\ref{solds}) into the
ansatz (\ref{dmetric1}) gives
\begin{equation}
dS_{(DS)}^{2}=dt^{2}-dr^{2}-a(r)(d\theta ^{2}+\sin ^{2}\theta d\varphi
^{2})-r_{0}^{2}e^{2\psi (r)}\left[ d\chi +\omega (r)dt+n_{[0]}\cos \theta
d\theta \right] ^{2}  \label{ansatz1}
\end{equation}
Up to a conformal factor and a rotation of the 5$^{th}$ coordinate this is
equivalent to the DS--solution for 5D wormhole/ flux tube configurations
\cite{dzhsin,ds} (more details can be found in refs. \cite{vsbd}, or \cite
{vs}). In (\ref{ansatz1}) $n_{[0]}$ is taken to be an integer; $r\in
\{-R_{0},+R_{0}\}$ ($R_{0}\leq $ $\infty $) and $r_{0}$ is a constant. All
functions $\psi (r)$ and $a(r)$ were taken to be even functions of $r$
satisfying $\psi ^{\prime }(0)=a^{\prime }(0)=0$. The coefficient
$\omega (r)$ in (\ref{ansatz1}) is treated as the $t$ --component of the
electromagnetic potential and $n_{[0]}\cos \theta $ as the $\theta $
-component. These electromagnetic potentials lead to the metric having
radial Kaluza-Klein ``electrical'' and ``magnetic'' fields. The 5D
Kaluza-Klein ``electric'' field is
\begin{equation}
E_{KK}=r_{0}\omega ^{\prime }e^{3\psi }=\frac{q_{0}}{a(r)}  \label{emf}
\end{equation}
the ``electric'' charge $q_{0}=r_{0}\omega ^{\prime }(0)$ can be
parametrized as $q_{0}=2\sqrt{a(0)}\sin \alpha _{0}$. The corresponding
dual, ``magnetic'' field is
\begin{equation}
H_{KK}=\frac{Q_{0}}{a(r)}  \label{hmf}
\end{equation}
with ``magnetic'' charge $Q_{0}=nr_{0}$ parametrized as $Q_{0}=2\sqrt{a(0)}
\cos \alpha _{0}$, The following circle relation
\begin{equation}
\frac{(q_{0}^{2}+Q_{0}^{2})}{4a(0)}{=1}  \label{emfm}
\end{equation}
relates the ``electric'' and ``magnetic'' charges. This relation between the
``electric'' and ``magnetic'' charges will still remain valid even when
these charges are generalized so as to have an anisotropic dependence on the
coordinates. In this case equation (\ref{emfm}) fixes the behavior of one
charge in terms of the other. For different relative values of the
``electric'' and ``magnetic'' charges it was found that the solution changed
from a wormhole (when the ``electric'' charged was larger than the
``magnetic'' charge) to a flux tube (when the ``magnetic'' charge was equal
to or larger than the ``electric'' charge).

\subsubsection{Warped, anisotropic solitonic extension of the DS--solution}

The simplest way to obtain anisotropic wormhole / flux tube solutions \cite
{vsbd} is to take $r_{0}^{2}$ from (\ref{ansatz1}) not as a constant, but as
``renormalized'' via $\,r_{0}^{2}\rightarrow \widehat{r}_{0}^{2}=\widehat{r}
_{0}^{2}(r,\theta ,v),$ where the anisotropic coordinate, $v$ could be $\chi
$ or $\varphi $ ({\it i.e.} we have two classes of anisotropic solutions,
one with the 5$^{th}$ coordinate and another with the
angular coordinate $\varphi ).$ A number of anisotropic wormhole/ flux tube
solutions in vacuum 5D gravity, for both cases -- anisotropies in $\chi $ or
$\varphi$ -- were constructed and analyzed in refs. \cite{vsbd,vs}. For
simplicity, in this paper we restrict our considerations only to warp
factors, anisotropies with respect to the extra dimensional
coordinate, and possible solitonic polarizations with respect to
the variables $\left( t,r,\theta ,\chi
\right) .$ Anisotropies on sets of coordinates like $\left( t,r,\theta
,\varphi \right) $ can be defined in a similar fashion and we omit such
considerations.

From the isotropic solution (\ref{solds}) we generated \cite{vsbd,vs} an
anisotropic solution by taking
\[
\widehat{h}_{4}(r,\theta )=h_{4}(r,\theta )=-a(r)\sin ^{2}\theta ,
\]
with $\eta _{4}=1$ so that $\widehat{h}_{4}^{\ast }=h_{4}^{\ast }=0,$ but $%
\widehat{h}_{5}^{\ast }=\eta _{5}^{\ast }(r,\theta ,\chi )h_{5}(r)\neq 0.$
We parametrized
\begin{equation}
\widehat{r}_{0}^{2}(r,\theta ,\chi )\simeq r_{0(0)}^{2}[1+\varpi (r,\theta
)\chi ]^{2}  \label{slin}
\end{equation}
so that
\[
\widehat{h}_{5}(r,\theta ,\chi )=\eta _{5}(r,\theta ,\chi )h_{5}(r),
\qquad 
h_{5}(r)=-r_{0}^{2}e^{2\psi (r)},
\qquad  \eta _{5}(r,\theta ,\chi )=[1+\varpi(r,\theta )\chi ]^{2}.
\]
The term $\varpi (r,\theta ) \chi$ parameterizes the dependence
of $\widehat{r}_{0}^{2}$ on $r, \theta , \chi$
In \cite{vsbd} we took the simple case of $\varpi (r,\theta)=const.$
in order to illustrate that the DS--solutions admit a running
of the ``electromagnetic'' constant with respect to the
5$^{th}$ coordinate $\chi $. In general, the physical constants
could also be anisotropically polarized through the holonomic
coordinates $(r,\theta),$ which is accomplished by letting
$\varpi =const. \rightarrow \varpi (r,\theta ).$

The ansatz functions $n_{2,3}$ depend on the
anisotropic variable $\chi $ in the following way
\[
n_{3}(r,\theta ,\chi )=n_{3[0]}(r,\theta )+n_{3[1]}(r,\theta )
[1+\varpi (r,\theta )\chi ]^{-2}.
\]
In the locally isotropic limit ($\varpi\chi \rightarrow 0$) 
one recovers the form of equation (\ref{ansatz1}) with
$n_{2[0,1]}=0$ , $n_{3[0]}=0$, $n_{3[1]}(r,\theta
)=n\cos \theta $ and $n_{1}=\omega (r).$

The 5D gravitational vacuum polarization induced by variation of the
``constant'' $\widehat{r}_{0}(\chi )$ renormalizes the electromagnetic
charge as $q(\theta ,\chi )=\widehat{r}_{0}(r=0,\theta ,\chi )\omega
^{\prime }(r=0)$. In terms of the angular parametrization the ``electric''
charge becomes
\[
q(\theta ,\chi )=2\sqrt{a(0)}\sin \alpha (\theta ,\chi ),
\]
and the ``electric'' field from (\ref{emf}) becomes
\[
E_{KK}=\frac{q(\theta ,\chi )}{a(r)}.
\]
The renormalization of the magnetic charge, $Q_{0}\rightarrow Q(\theta ,\chi
)$, can be obtained using the renormalized ``electric'' charge in
relationship (\ref{emfm}) and solving for $Q(\theta ,\chi ).$ The form of
(\ref{emfm}) implies that the running of $Q(\theta ,\chi )$ will be the
opposite that of $q(\theta ,\chi )$. For example, if $q(\theta ,\chi )$
increases with $\chi $ then $Q(\theta ,\chi )$ will decrease. The locally
anisotropic polarizations $\alpha (\theta ,\chi )$ are either defined from
experimental data or computed from a quantum model of 5D gravity.

Thus the following ansatz functions for (\ref{dmetric1})
\begin{eqnarray}
g_{1} &=&1,\qquad g_{2}=-1,\qquad g_{3}=-a(r) , \qquad
k=k_{t}=k_{r}=k_{\theta }=0,
\label{set2} \\
\widehat{h}_{4} &=&h_{4}=-a(r)\sin ^{2}\theta ,\qquad \eta _{4}=1  \nonumber
\\
\widehat{h}_{5} &=&\eta _{5}h_{5},\qquad h_{5}(r)=-r_{0}^{2}e^{2\psi
(r)},\qquad \eta _{5}=[1+\varpi (r,\theta )\chi ]^{2}  \nonumber \\
n_{1}&=& \omega (r),\qquad n_{2}=0,\qquad n_{3}=n_{[0]}\cos
\theta \lbrack 1+\varpi (r,\theta )\chi ]^{-2},  \nonumber
\end{eqnarray}
generates an anisotropic wormhole / flux tube metric whose electromagnetic
charges have become dependent on $\left( r, \theta , \chi \right)$.

Next we further modify the locally anisotropic configuration in equation
(\ref{set2}) to include a warp factor and a solitonic deformation.
Instead of $\widehat{h}_{5}$ from (\ref{set2}) we choose
\begin{equation}
\widehat{h}_{5}=h_{5}^{[KP,SG]}\left( r,\theta ,\chi \right) =\eta
_{5}^{[KP,SG]}\left( r,\theta ,\chi \right) \eta _{5}\left( r,\theta \right)
h_{5}(r),\qquad h_{5}(r)=-r_{0}^{2}e^{2\psi (r)},  \label{auxx1}
\end{equation}
where $\eta _{5}^{[KP,SG]}\left( r,\theta ,\chi \right) $ is defined
by the requirement that $\widehat{h}_{5}$ be a solution of the KdP
equation (\ref{kdp}), or of the SG equation (\ref{sineq}). Then
we modify $\widehat{h}_{4}$ as
\begin{equation}
\widehat{h}_{4}\left( r,\theta ,\chi \right) =\eta _{4}^{[KP,SG]}\left(
r,\theta ,\chi \right) h_{4}(r,\theta ),  \label{auxx2}
\end{equation}
where $h_{4}=-a(r)\sin ^{2}\theta $ as in (\ref{set2}). Remembering
that $h_{4}(r,\theta ) = h_{[0]}^{2}\eta _{5}\left(r,\theta \right)
h_{5}(r)$ with $h_{[0]}=const$, then gives  $\eta
_{4}^{[KP,SG]}\left( r,\theta ,\chi \right)$ from equation
(\ref{p1b}) as
\[
\eta _{4}^{KP,SG}=[(\sqrt{\eta _{5}^{KP,SG}})^{^{\ast }}]^{2}.
\]
The values of $n_{i}=n_{i}^{[KP,SG]}$ are computed by introducing
$\widehat{h}_{4}$ and $\widehat{h}_{5}$ into the formulas (\ref{n}).
Next we require that $n_{i[0,1]}(x_i)$ satisfy boundary conditions so as
to give nontrivial values for $n_{1}^{[KP]}=\omega (r)$ and
$n_{3}=n_{[0]}\cos \theta \ \mu^{\lbrack KP,SG]}\left( r,\theta ,
\chi \right) ,$ where  $\mu ^{\lbrack KP,SG]}\left( r,\theta ,\chi \right)$
can be picked out after performing the  $\chi$ integration in
equation (\ref{n}) for the functions from equations (\ref{auxx1})
and (\ref{auxx2}).

Finally we introduce a conformal factor (\ref{expconf}). This results in a
constant shift of the second order anisotropy $\zeta _{1}\left( t,\chi
\right) =|\chi |\partial _{t}\ln |k\left( t\right) |, \zeta _{2}=
\zeta _{3}=0.$ Collecting all the results together we see that
\begin{eqnarray}
g_{1} &=&1,\qquad g_{2}=-1,\qquad g_{3}=-a(r),\qquad
\Omega =e^{-k\left( t\right) |\chi |},
\label{set3} \\
\widehat{h}_{4} &=&\eta _{4}h_{4} , \qquad h_{4}=-a(r)\sin ^{2}\theta ,
\qquad \eta_{4}^{KP,SG}=[(\sqrt{\eta _{5}^{KP,SG}})^{^{\ast }}]^{2},
\nonumber \\
\widehat{h}_{5} &=&\eta _{5}^{[KP,SG]}\left( r,\theta ,\chi \right) \eta
_{5}(r,\theta )h_{5} , \qquad h_{5}(r)=-r_{0}^{2}e^{2\psi (r)} ,
\qquad h_{4}(r,\theta )=h_{[0]}^{2}\eta_{5}\left( r,\theta \right) h_{5}(r),  \nonumber \\
\zeta _{1} &=& |\chi |\partial _{t}\ln |k\left( t\right)
|, \qquad n_{1}=\omega (r), \qquad n_{2}=0, \qquad
n_{3}=n_{[0]}\cos \theta \mu ^{[KP,SG]}
\left( r,\theta ,\chi \right) ,  \nonumber
\end{eqnarray}
gives a new, exact solution of the 5D vacuum Einstein equations. It has an
exponential warp factor in the  extra dimensional coordinate, and is also
time dependent. The isotropic wormhole / flux tube
configuration of ref. \cite{ds} has been self--consistently
embedded on an anisotropic, KdP or SG solitonic gravitational background.
The ``electromagnetic'' constants of the solutions are anisotropically
renormalized and scale with respect to the 5$^{th}$ coordinate.
The Kaluza--Klein magnetic field component exhibits a 3D solitonic
polarization.

The final remark in this section is that we can also consider 3D solitons
of the KdP or SG equations with a time like dependence in the
polarization functions like $\eta _{4,5}^{[KP,SG]}(t,\theta ,\chi ).$
In this case the solution would describe a warped anisotropic
wormhole / flux tube configuration moving self-consistently in an
effective 4D spacetime. Such a configuration could be thought of
as a traveling gravitational soliton which might be searched for
in a gravitational wave detector such as LIGO.

\section{Gravitational Polarization of Kaluza-Klein Charges}

We can further generalize the form (\ref{set2}) and (\ref{set3}) to generate
new solutions of the 5D vacuum Einstein equations with deformations of the
constants $r_{0}^{2}$ and $n_{[0]}$ with respect to the $\theta $ variable.
These $\theta $ deformations take the form of the equation for an ellipsoid
in polar coordinates. This again leads to renormalization of electric ($q$)
and magnetic ($Q$) charges.

\subsection{Gravitational renormalization of Kaluza-Klein charges via
variable $r_0$}

In this subsection we give a solution for which the Kaluza-Klein charges are
gravitationally renormalized by the radius becoming dependent on $\theta$
({\it i.e.} in equation (\ref{emfm}) $a(0) \rightarrow a(\theta )$.

The easiest way to obtain such $\theta $--polarizations for the solutions of
(\ref{slin}) is to recast the ansatz functions (\ref{set3}) as
$\widehat{h}_{5}=\overline{\eta }_{5}h_{5}(r)$ with
\begin{equation}
\overline{\eta }_{5}=\left[ 1+\varepsilon _{r}\cos \theta \right] ^{-2}\eta
_{5}^{[KP,SG]}\left( r,\theta ,\chi \right) \eta _{5}(r,\theta ),
\label{eta1a}
\end{equation}
and
\[
\widehat{r}_{0}^{2}(r,\theta ,\chi )\simeq r_{0(0)}^{2}\left[ 1+\varepsilon
_{r}\cos \theta \right] ^{-2}\eta _{5}(r,\theta )\eta _{5}^{[KP,SG]}\left(
r,\theta ,\chi \right) ,
\]
where $\varepsilon _{r}$ is the eccentricity of an ellipse. The
``constant'', $\widehat{r}_{0}$, has both an elliptic variation in $\theta $
({\it i.e.} $r_{0(0)}\left[ 1+\varepsilon _{r}\cos \theta \right] ^{-1}$)
and a solitonic variation with respect to the 5$^{th}$ coordinate, $\chi$,
and the variables $r,\theta$.

With these solutions the 5D Kaluza-Klein charges
get renormalized through the elliptic variation of
$\widehat{r}_{0}(r,\theta ,\varepsilon _{r},\chi )$, and the
``electric'' charge becomes
\[
q(r,\theta ,\chi )=r_{0}\sqrt{\overline{\eta }_{5}(r,\theta ,\varepsilon
_{r},\chi )}\omega ^{\prime }(0)=\sqrt{\overline{\eta }_{5}(r,\theta
,\varepsilon _{r},\chi )}q_{0}
\]
In terms of the angular parametrization we find
\[
q(r,\theta ,\chi )=2\sqrt{a(0)}\sqrt{\overline{\eta }_{5}(r,\theta
,\varepsilon _{r},\chi )}\sin \alpha _{0},
\]
The ``electric'' field (\ref{emf}) transforms into
\[
E_{KK}=\frac{q(r,\theta ,\chi )}{a(r)}=\frac{q_{0}}{(\sqrt{\overline{\eta }
_{5}})^{-1}a(r)}
\]
$(\sqrt{\overline{\eta }_{5}})^{-1}$ can be treated as an anisotropic,
gravitationally induced permittivity. The renormalization of the magnetic
charge, $Q_{0}\rightarrow Q(r,\theta ,\chi )$, can be obtained from
equation (\ref{emfm}) using
$q(r,\theta ,\chi )$ from above. In this case the
corresponding dual, ``magnetic'' field is $H_{KK}=Q(r,\theta ,\chi )/a(r)$
with the ``magnetic'' charge $Q_{0}=n_{[0]}r_{0}$ given by
\[
Q=2\sqrt{a(0)}\sqrt{\overline{\eta }_{5}(r,\theta ,\varepsilon _{r},\chi )}
\cos \alpha _{0},
\]
These gravitationally polarized charges satisfy the circumference equation
(\ref{emfm}) with variable radius $2\sqrt{a(0)}\sqrt{\overline{\eta }
_{5}(r,\theta ,\varepsilon _{r},\chi )}$
\begin{equation}
\frac{(q_{0}^{2}+Q_{0}^{2})}{4a(0)\overline{\eta }_{5}(r,\theta ,\varepsilon
_{r},\chi )}{=1.}  \label{emfm1}
\end{equation}

The new result -- as compared with the elliptically polarized solutions
of \cite{vs} -- is that we have an additional
dependence on the radial coordinate $r$. Also the anisotropic,
gravitationally induced permittivity, $(\sqrt{\overline{\eta }_{5}})^{-1}$,
arises in connection with the 3D KdP or SG soliton solutions which
have been embedded in the 5D Einstein equations.

\subsection{Gravitational renormalization of Kaluza-Klein charges via $r_0$
and $n$}

A different class of solutions from those given in
equations (\ref{set2}) and
(\ref{set3}) can be constructed if, in addition to $r_{0}$, we allow the
$n_{[0]}$ in the $n_{[0]}\cos \theta $ term from
equation (\ref{ansatz1}) to vary. For solutions
with anisotropic extra dimensional coordinate, this variable
$n_{[0]}$ will affect $n_{3}$. The variability of $r_{0}$ and $n_{[0]}$ is
parameterized using the gravitational vacuum polarizations $\kappa
_{r}\left( r,\theta ,\chi \right) $ and $\kappa _{n}\left( r,\theta ,\chi
\right) $ as
\[
r_{0}\rightarrow \widehat{r}_{0}=\frac{r_{0}}{\kappa _{r}\left( r,\theta ,
\chi\right)} \qquad \mbox{and } \qquad
n_3 \rightarrow \widehat{n}_3 =
\frac{n_{[0]}}{\kappa _{n}\left( r,\theta ,\chi
\right)}
\]
where
\begin{equation}
\kappa _{r}\left( r,\theta ,\chi \right) =[\sqrt{\eta _{4}\left( r,\theta
,\chi \right) }]^{-1} \qquad \mbox{or } \qquad
=[\sqrt{\eta _{5}\left( r,\theta,\chi \right) }]^{-1}.  \label{eta2a}
\end{equation}
The polarized charges are
\[
q=\frac{q_{0}}{\kappa _{r}}=\frac{2\sqrt{a(0)}\sin \alpha _{0}}{\kappa _{r}}
\]
and
\[
Q=\frac{Q_{0}}{\kappa _{n}}=\frac{2\sqrt{a(0)}\cos \alpha _{0}}{\kappa _{n}},
\]
Using these charges in equation (\ref{emfm}) gives
the formula for an ellipse in
the charge space coordinates $\left( q_{0},Q_{0}\right) ,$
\begin{equation}
\frac{q_{0}^{2}}{4a(0)\kappa _{r}^{2}}
+\frac{Q_{0}^{2}}{4a(0)\kappa _{n}^{2}}{=1,}  \label{emfm3}
\end{equation}
the axes of the ellipse are $2\sqrt{a(0)}\kappa _{r}$ and $2\sqrt{a(0)}
\kappa _{n}.$ Equation (\ref{emfm3}) contains as a particular case 
equation (\ref{emfm1}). A new result for such ellipsoidal polarizations
(compared with a similar solution from \cite{vs}) is that the parameters of
the ellipse (the Kaluza-Klein charges) are modified by an additional
solitonic dependence. This describes  3D solitonic waves in the charge
space induced by the 5D vacuum gravitational interactions.

\section{Warped Wormholes in Various Non-spherical Backgrounds}

The locally anisotropic wormhole/flux tube solutions presented in the
previous sections are anisotropic deformations from a spherical 3D
hypersurface background. These solutions can be generalized to other
rotational hypersurface geometry backgrounds. In ref. \cite{vs} we gave the
explicit forms for such generalized solutions and analyzed their basic
properties. The aim of this section is to show that these anisotropic
wormhole solutions with ellipsoidal, cylindrical, bipolar
and toroidal backgrounds can be solitonically deformed and
have a warp factor. The notation and metric relations for the 3D
Euclidean rotational hypersurfaces that we use will be those of ref. \cite
{korn}. In the appendix we outline the necessary results for 3D rotation
hypersurfaces and explain the notation (see also refs. \cite{v}).

In order to construct wormholes which exhibit the various 3D geometries
cataloged in the appendix, we will associate respectively one of the ansatz
functions of the wormhole solutions with the metric
component, $g_{ss} (x^{2},x^{3})$, from (\ref
{hsuf1a}), (\ref{hsuf1b}), (\ref{mbipcy}) and (\ref{mtor}).
For the solutions with anisotropic dependence on the extra
dimension coordinate this is accomplished by letting
$h_{4}=g_{ss}(x^{2},x^{3})$ and then solitonically deforming
$h_{4}\rightarrow \widehat{h}_{4}$ and $h_{5}\rightarrow \widehat{h}_{5}.$
In each case $h_{4,5}$ is multiplied by the
corresponding gravitational polarizations, $\eta _{4,5}$, so as to give
wormhole/flux tube configurations of the form (\ref{set2}) and (\ref{set3}),
or their modifications presented in section IV. In the previous
sections the coordinates were parametrized as $x^{2}=r,$
$x^{3}=\theta $ and $y^{5}=\varphi $. In this section the set
$(x^{2},x^{3},y^{5})$ are spatial coordinates of a 3D rotation
hypersurface. The time like and extra dimensional coordinates are the same
as in the previous sections, {\it i.e.} $x^{1}=t$ and $y^{4}=\chi .$

Straightforward calculations show that for the five 3D geometries
given in the appendix, the metrics defined by (\ref{set2}) or (\ref{set3})
are generalized to a new class of exact solutions of the 5D vacuum
Einstein equations with the coordinates given by
\begin{eqnarray}
x^{k} &=&\left\{
\begin{array}{l}
(t,u,v),0\leq u<\infty ,\ 0\leq v\leq \pi ,\cosh u\geq 1,
\mbox{ ellipsoid
(\ref{relhor})}; \\
(t,u,v),0\leq u<\infty ,\ 0\leq v\leq \pi ,\sinh u\geq 0,
\mbox{ ellipsoid
(\ref{relhor1})}; \\
(t,u,v),0\leq u<\infty ,\ 0\leq v\leq \pi ,\cosh u\geq 1,
\mbox{cylinder
(\ref{elcyl})}; \\
(t,\tau ,\xi ),-\infty <\tau <\infty ,0\leq \xi <\pi ,
\mbox{bipolar
(\ref{bip})}; \\
(t,\tau ,\xi ),0\leq \tau <\infty ,-\pi <\xi <\pi ,
\mbox{torus
(\ref{torus})};
\end{array}
\right.  \nonumber \\
y^{4} &=&s=\chi ,\qquad y^{5}=p=\left\{
\begin{array}{l}
\varphi \in \lbrack 0,2\pi ),\mbox{ ellipsoids };\mbox{bipolar };
\mbox{torus}; \\
z\in (-\infty ,\infty ),\mbox{cylinder};
\end{array}
\right.  \nonumber
\end{eqnarray}
and the ansatz functions given by
\begin{eqnarray}
g_{1} &=&1,\ g_{2}=-1,\ g_{3}=-1,\ \Omega =e^{-k\left( t\right) |\chi |},
\label{ellipschi} \\
\widehat{h}_{4} &=&\eta _{4}h_{4},\qquad h_{4}=g_{ss}(x^{2},x^{3})=\left\{
\begin{array}{l}
\frac{\sinh ^{2}u\sin ^{2}v}{\sinh ^{2}u+\sin ^{2}v},%
\mbox{ ellipsoid
(\ref{hsuf1a})}; \\
\frac{\sinh ^{2}u\cos ^{2}v}{\sinh ^{2}u+\cos ^{2}v},%
\mbox{ellipsoid
(\ref{hsuf1b})}; \\
\frac{\rho ^{-2}(u,v)}{\sinh ^{2}u+\sin ^{2}v},\mbox{cylinder (\ref{elcyl})};
\\
\sin ^{2}\xi ,\mbox{bipolar  (\ref{bip})};\mbox{torus  (\ref{torus})};
\end{array}
\right. ,  \nonumber \\
\eta _{5} &=&\left\{
\begin{array}{l}
\mu ^{\lbrack KP,GS]}\left( x^{2},x^{3},\chi \right) ,
\mbox{ see
(\ref{set2})}; \\
\left[ 1+\varepsilon _{r}\cos \theta \right] ^{-2}\mu ^{\lbrack
KP,GS]}\left( x^{2},x^{3},\chi \right) ^{2},\mbox{
see  (\ref{eta1a})}; \\
1/\kappa _{r}^{2}(x^{2},x^{3},\chi ),\mbox{ see  (\ref{eta2a})};
\end{array}
\right. ,  \nonumber \\
\widehat{h}_{5} &=&\eta _{5}h_{5},\; \;
h_{5}(x^{2},x^{3},\chi )=-r_{0}^{2}\exp
\{2\psi \lbrack (x^{2},x^{3},\chi )]\};\ ,\; \; \zeta _{1}=\zeta
_{1}=|\chi |\partial _{t}\ln |k\left( t\right) |;  \nonumber \\
r &=&\widetilde{a}^{(invers)}\left( x^{2},x^{3},\chi \right) \mbox{ from }
\mbox{
(\ref{relhor})};\mbox{ (\ref{relhor1})};\mbox{ (\ref{elcyl})};
\mbox{
(\ref{bip})};\mbox{(\ref{torus})};  \nonumber \\
\ n_{1} &=&\omega (x^{2}),\qquad n_{2}=0,\qquad n_{3}=n\cos x^{3}\times
\left\{
\begin{array}{l}
\mu ^{\lbrack KP,GS]}\left( x^{2},x^{3},\chi \right) ,
\mbox{ see
(\ref{set2})}; \\
\mu ^{\lbrack KP,GS]}\left( x^{2},x^{3},\chi \right) ,
\mbox{ see
(\ref{set3})}; \\
1/\kappa _{n}(x^{2},x^{3},\chi ),\mbox{ see  (\ref{emfm3})}.
\end{array}
\right. ;  \nonumber
\end{eqnarray}
where the function $\widehat{h}_{5}$ is a solitonic solution of
the KdP or SG equations.

Equations (\ref{ellipschi}) describe warped, anisotropic wormhole / flux
tube, solitonic configurations which are defined self--consistently in the
various rotational hypersurface backgrounds listed above. As in the case of
the spherical background these solutions have an anisotropic deformation
with respect to the given hypersurface backgrounds.

\section{Conclusions}

Some recent work in modern string theory, extra dimensional gravity
and particle physics is based on the idea
that our 4D universe is embedded in some higher dimension spacetime,
but with movement into the higher dimensions being
suppressed by an exponential warp factor \cite{rs}. The construction and
study of wormhole/flux tube solutions in these RS-type
models is of fundamental importance in understanding these
theories, especially their non-perturbative aspects. Such solutions are
difficult to find, and the solutions which are known usually have a high
degree of symmetry. In this paper we have applied the method of anholonomic
frames to construct warped, solitonic wormholes and flux
tubes in 5D Kaluza-Klein theory. These solutions have local anisotropy which
would make their study using holonomic frames difficult. This 
demonstrates the usefulness of the anholonomic frames
method in handling anisotropic solutions.
Most physical situations do not possess a high degree
of symmetry, and so the anholonomic frames method provides a useful
mathematical framework for studying these less symmetric configurations.

The main result of our work is that we can construct exact solutions in 5D
gravity with a warp factor in the extra dimension coordinate which is
induced not from any brane energy--momentum configurations but by a specific
type of second order anisotropy of the bulk vacuum gravity.

The second key result is the demonstration that off--diagonal metrics in 5D
Kaluza--Klein theory can be parametrized into forms that define new,
interesting classes of solutions of Einstein's vacuum equations. These
solutions represent wormhole and flux tube configurations which are locally
anisotropic. These anisotropic solutions reduce to previously known
spherically symmetric wormhole metrics \cite{chodos,dzhsin,ds} in the local
isotropic limit. These anisotropic solutions also extend the idea of Salam,
Strathee and Percacci \cite{sal} that including off--diagonal components in
higher dimensional metrics gives rise to gauge fields and charges. Not only
do we find ``electric'' and``magnetic'' charges for our solutions, but the
anisotropies in the 5$^{th}$ coordinate ($\chi$) and/or in the angular
coordinate ($\varphi$) give a gravitational scaling or running of these
Kaluza-Klein charges. Such a gravitational scaling of charges could provide
an experimental signature for the presence of extra dimensions ({\it i.e.}
if some charge were observed to exhibit a running which was not in agreement
with that given by 4D quantum field theory this could be evidence for a
gravitational running of the charge).

In connection to the status of off--diagonal metrics in 5D gravity we note
that in refs. \cite{vgg} it was shown that the class of metrics (\ref
{cmetric}) satisfying vacuum Einstein equations (\ref{ricci1a})--(\ref
{confeq}) contains as a particular case solutions where the Schwarzschild
potential $\Phi =-M/(M_{{\rm p}}^{2}r)$, ($M_{{\rm p}}$ is the
effective Planck mass on the brane) is modified to
\[
\Phi =-{\frac{M\sigma _{m}}{M_{{\rm p}}^{2}r}}+{\frac{Q\sigma _{q}}{2r^{2}}}
\,,
\]
where the ``tidal charge'' parameter $Q$ may be positive or negative. This
points to the possibility of anisotropically modifying Newton's law via
an effective anisotropic masses $M\sigma _{m}(t,r,\theta ),$
and with an effective gravitational electric-like charge $Q\sigma
_{q}(t,r,\theta )$. For diagonal metrics the effective polarizations are
$\sigma _{m}=\sigma _{q}=1.$ The off--diagonal metrics and anholonomic
frames in extra dimensional vacuum gravity can be used to give
non--trivial reductions and trappings to lower dimensional models of
wormhole and black hole physics. This can result in anisotropic
polarizations and scaling of the parameters of the solution,
as well as modifications of Newtonian gravity.

In the first part of this paper the warped anisotropic solutions were
constructed as deformations from a spherical background. We considered 3D
solitonic deformations (Kadomtsev--Petviashvili and sine--Gordon equations).
In the final section of this paper we showed that it is possible to
construct a large variety of such warped anisotropic solutions as solitonic
deformations from various background geometries: elliptic (elongated and
flattened), cylindrical, toroidal and bipolar.

\section*{Acknowledgments}

D.S. would like to thank V. Dzhunushaliev for discussions related to this
work. S. V. work is supported both by a 2000--2001 California State
University Legislative Award and a NATO/Portugal fellowship grant at the
Technical University of Lisabon.

\appendix

\section{3D Rotation Hypersurfaces}

In this appendix we collect together the basic 3D geometric results \cite
{v,korn} used in section {\bf V}.

\subsection{Elongated rotation ellipsoid hypersurfaces}

An elongated rotation ellipsoid hypersurface (a 3D e--ellipsoid) is given by
the formula
\begin{equation}
\frac{x^{2}+y^{2}}{\sigma ^{2}-1}+\frac{z^{2}}{\sigma ^{2}}=\widetilde{a}%
^{2}(r),  \label{relhor}
\end{equation}
where $\sigma \geq 1,$ and $x,y,z$ are the usual Cartesian coordinates.
$\widetilde{a}(r)$ is similar to the radius in the spherical symmetric case.
The 3D, ellipsoidal coordinate system is defined
\[
x=\widetilde{a}\sinh u\sin v\cos s,\qquad y=\widetilde{a}\sinh u\sin v\sin
s,\qquad z=\widetilde{a}\ \cosh u\cos v,
\]
where $\sigma =\cosh u$ and $0\leq u<\infty ,\ 0\leq v\leq \pi ,\ 0\leq
s<2\pi .$ The hypersurface metric is
\begin{equation}
g_{uu}=g_{vv}=\widetilde{a}^{2}\left( \sinh ^{2}u+\sin ^{2}v\right) ,\qquad
\widetilde{g}_{ss}=\widetilde{a}^{2}\sinh ^{2}u\sin ^{2}v.  \label{hsuf1}
\end{equation}
It will be more useful to consider a conformally transformed metric, where
the components in equation (\ref{hsuf1})
are multiplied by the conformal factor
$\widetilde{a}^{-2}\left( \sinh ^{2}u+\sin ^{2}v\right) ^{-1},$ giving
\begin{eqnarray}
ds_{(3e)}^{2} &=&du^{2}+dv^{2}+g_{ss}(u,v)ds^{2}  \label{hsuf1a} \\
g_{ss}(u,v) &=&\sinh ^{2}u\sin ^{2}v/(\sinh ^{2}u+\sin ^{2}v).  \nonumber
\end{eqnarray}

\subsection{Flattened rotation ellipsoid hypersurfaces}

In a similar fashion we consider the hypersurface equation for a flattened
rotation ellipsoid (a 3D f--ellipsoid),
\begin{equation}
\frac{x^{2}+y^{2}}{1+\sigma ^{2}}+\frac{z^{2}}{\sigma ^{2}}= \widetilde{a}
^{2}(r),  \label{relhor1}
\end{equation}
here $\sigma \geq 0$ and $\sigma =\sinh u.$ In this case the 3D coordinate
system is defined as
\[
x=\widetilde{a}\cosh u\sin v\cos s,\qquad y=\widetilde{a}\cosh u\sin v\sin
s,\qquad z=\widetilde{a}\sinh u\cos v,
\]
where $0\leq u<\infty ,\ 0\leq v\leq \pi ,\ 0\leq s<2\pi .$ The hypersurface
metric is
\begin{equation}
g_{uu}=g_{vv}=\widetilde{a}^{2}\left( \sinh ^{2}u+\cos ^{2}v\right) \qquad
g_{\varphi \varphi }=\widetilde{a}^{2}\sinh ^{2}u\cos ^{2}v,  \nonumber
\end{equation}
Again we consider a conformally transformed version of this metric
\begin{eqnarray}
ds_{(3f)}^{3} &=&du^{2}+dv^{2}+g_{ss}(u,v)ds^{2},  \label{hsuf1b} \\
g_{ss}(u,v) &=&\sinh ^{2}u\cos ^{2}v/(\sinh ^{2}u+\cos ^{2}v).  \nonumber
\end{eqnarray}

\subsection{Ellipsoidal cylindrical hypersurfaces}

The formula for an ellipsoidal cylindrical hypersurface is
\begin{equation}
\frac{x^{2}}{\sigma ^{2}}+\frac{y^{2}}{\sigma ^{2}-1}=\rho ^{2},\ z=s,
\label{elcyl}
\end{equation}
where $\sigma \geq 1.$ The 3D radial coordinate is given as $\widetilde{a}
^{2}=\rho ^{2}+s^{2}.$ The 3D coordinate system is defined
\[
x=\rho \cosh u\cos v,\qquad y=\rho \sinh u\sin v,\qquad z=s,
\]
where $\sigma =\cosh u$ and $0\leq u<\infty ,\ 0\leq v\leq \pi $. Using the
expressions for $x,y$ and equation (\ref{elcyl}) we can make the
change $\rho(x,y)\rightarrow \rho (u,v)$. The hypersurface metric is
\[
g_{uu}=g_{vv}=\rho ^{2}(u,v)\ \left( \sinh ^{2}u+\sin ^{2}v\right) ,\qquad
g_{ss}=1;
\]
we will again consider a conformally transformed version of this metric
\begin{eqnarray}
ds_{(3c)}^{2} &=&du^{2}+dv^{2}+g_{ss}(u,v,\rho (u,v))ds^{2},  \label{cylin1}
\\
g_{ss}(u,v) &=&1/\rho ^{2}(u,v)\ (\sinh ^{2}u+\sin ^{2}v).  \nonumber
\end{eqnarray}

\subsection{ Bipolar coordinates}

Now we consider a bipolar hypersurface given by the formula
\begin{equation}
\left( \sqrt{x^{2}+y^{2}}-\frac{\widetilde{a}(r)}{\tan \xi }\ \right)
^{2}+z^{2}=\frac{\widetilde{a}^{2}(r)}{\sin ^{2}\xi },  \label{bip}
\end{equation}
which describes a hypersurface obtained by rotating the circles
\[
\left( y-\frac{\widetilde{a}(r)}{\tan \xi }\right) ^{2}+z^{2} =\frac{
\widetilde{a}^{2}(r)}{\sin ^{2}\xi }
\]
around the $z$ axis; because $|\tan \xi |^{-1}<|\sin \xi |^{-1},$ the
circles intersect the $z$ axis. The relationship between the Cartesian
coordinates and the bipolar coordinates is
\[
x=\frac{\widetilde{a}(r)\sin \xi \cos s}{\cosh \tau -\cos \xi },\qquad y=
\frac{\widetilde{a}(r)\sin \xi \sin s}{\cosh \tau -\cos \xi },\qquad z=\frac{
\widetilde{a}(r)\sinh \tau }{\cosh \tau -\cos \xi },
\]
where $-\infty <\tau <\infty ,0\leq \xi <\pi ,0\leq s<2\pi $. The
hypersurface metric is
\[
g_{\tau \tau }=g_{\xi \xi }=\frac{\widetilde{a}^{2}(r)}{\left( \cosh \tau
-\cos \xi \right) ^{2}},\qquad g_{ss}=\frac{\widetilde{a}^{2}(r)\sin ^{2}\xi
}{\left( \cosh \tau -\cos \xi \right) ^{2}},
\]
which, after multiplication by the conformal factor $\left( \cosh \tau -\cos
\sigma \right) ^{2}/\rho ^{2}$ becomes
\begin{equation}
ds_{(3b)}^{2}=d\tau ^{2}+d\xi ^{2}+g_{ss}(\xi )ds^{2},\qquad g_{ss}(\xi
)=\sin ^{2}\xi .  \label{mbipcy}
\end{equation}

\subsection{Toroidal coordinates}

Now we consider a toroidal hypersurface with nontrivial topology given by
the formula
\begin{equation}
\left( \sqrt{x^{2}+y^{2}}-\widetilde{a}(r)\ (\coth \xi )\right) ^{2}+ z^{2}=
\frac{\widetilde{a}^{2}(r)}{\sinh ^{2}\xi },  \label{torus}
\end{equation}
the relationship to the Cartesian coordinates is given by
\[
x=\frac{\widetilde{a}(r)\sinh \tau \cos s}{\cosh \tau -\cos \xi },\qquad y=
\frac{\widetilde{a}(r)\sin \xi \sin s}{\cosh \tau -\cos \xi },\qquad z=\frac{
\widetilde{a}(r)\sinh \xi }{\cosh \tau -\cos \xi },
\]
where $-\pi <\xi <\pi ,0\leq \tau <\infty ,0\leq s<2\pi $. The hypersurface
metric is
\[
g_{\sigma \sigma }=g_{\tau \tau }=\frac{\widetilde{a}^{2}(r)}{\left( \cosh
\tau -\cos \xi \right) ^{2}},\qquad g_{ss}=\frac{\widetilde{a}^{2}(r)\sin
^{2}\xi }{\left( \cosh \tau -\cos \xi \right) ^{2}}
\]
After multiplication by the conformal factor $\left( \cosh \tau -\cos \sigma
\right) ^{2}/\widetilde{a}^{2}(r)$ this takes the same form as (\ref{mbipcy}
)
\begin{equation}
ds_{(3t)}^{3}=d\tau ^{2}+d\xi ^{2}+g_{ss}(\xi )ds^{2},\qquad g_{ss}(\xi
)=\sin ^{2}\xi ,  \label{mtor}
\end{equation}
Although this looks identical to the metric in (\ref{mbipcy}) the
coordinates $\left( \tau ,\xi ,s\right) $ have different meanings in each
case. This can be seen by the different ranges for the two cases.


\begin{references}
\bibitem{v}  S. Vacaru, \ JHEP {\bf 04}: 009 (2001);\ Ann. Phys. (NY) {\bf %
290}, 83 (2001); S. Vacaru and F. C. Popa, Class. Quant. Grav. {\bf 18}, 1
(2001).

\bibitem{vst}  S. Vacaru, Ann. Phys. (NY) {\bf 256}, 39 (1997);\ Nucl. Phys.
{\bf B434}, 590 (1997);\ J. Math. Phys. {\bf 37}, 508 (1996);\ JHEP {\bf 09}
: 011 (1998);\ Phys. Lett. {\bf B 498,} 74 (2001).

\bibitem{vsbd}  S. Vacaru, D. Singleton, V. Botan and D.\ Dotenco, Phys.
Lett. {\bf B 519,} 249 (2001).

\bibitem{vtheorem}  S.\ Vacaru, A New Metod of Constructing Black Hole
Solutions in Einstein and 5D Gravity, hep-th/0110250.

\bibitem{mor}  M. S. Morris and K. S. Thorne, Am. J. Phys., {\bf 56}, 395
(1988);\ S. Giddings and A. Strominger, Nucl.Phys., {\bf B306}, 890 (1988);\
K. A. Bronnikov and J. C. Fabris, Grav and Cosmol. {\bf 3}, 67 (1997);\ M.
Rainer and A. Zhuk, Phys.Rev., {\bf D54}, 6186 (1996);\ M. Visser, {\it 
Lorentzian Wormholes: from Einstein to Hawking}, (AIP, New York, 1995).

\bibitem{sal}  A. Salam and J. Strathdee, Ann. Phys. (NY) {\bf 141}, 316
(1982);\ R. Percacci, J. Math. Phys. {\bf 24}, 807 (1983).

\bibitem{chodos}  A. Chodos and S. Detweiler, Gen. Rel. Grav. {\bf 14}, 879
(1982);\ G. Cl\'{e}ment, Gen. Rel. Grav. {\bf 16}, 131 (1984).\

\bibitem{dzhsin}  V. D. Dzhunushaliev, Izv. Vuzov, ser. Fizika, {\bf 78,} N6
(1993)( in Russian);\ Grav. and Cosmol. {\bf 3}, 240 (1997);\ V. D.
Dzhunushaliev, Gen. Rel. Grav. {\bf 30}, 583 (1998);\ V. D. Dzhunushaliev,
Mod. Phys. Lett. A {\bf 13}, 2179 (1998);\ V. D. Dzhunushaliev and D.
Singleton, Class. Quant. Grav. {\bf 16}, 973 (1999).

\bibitem{ds}  V. D. Dzhunushaliev and D. Singleton, Phys. Rev., {\bf D59},
064018 (1999).

\bibitem{vs}  S. Vacaru and D. Singleton, Ellipsoidal, Cylindrical, Bipolar
and Toroidal Wormholes, hep-th/0110272

\bibitem{rs}  L. Randall and R. Sundrum, Phys. Rev. Lett. {\bf 83} 3370
(1999); Phys. Rev. Lett. {\bf 83} 4690 (1999).

\bibitem{vgg}  S. Vacaru, Off--Diagonal 5D Metrics and Mass Hierarchies with
Anisotropies and Running Constants, hep-ph/ 0106268; \ S. Vacaru and E.
Gaburov, \ Anisotropic Black Holes in Einstein and Brane Gravity, hep-th/
0108065; S.\ Vacaru and D. Gontsa, Off--Diagonal Metrics and Anisotropic
Brane Inflation, hep--th/ 0109114.

\bibitem{radion}  Nima Arkani--Hamed, S. Dimopoulos, N. Kaloper and J.
March--Russell, Nucl. Phys. {\bf D567}, 189 (2000); C.\ Csaki, M. Graesser,
J. Terning, Phys. Lett. {\bf B456,} 16 (1999); \ J. Cline, Phys. Rev., {\bf %
D61}, 023513 (2000); E. Halyo, JHEP {\bf 9909} : 012 (1999); E. Flanagnan,
S. H. Henry Tye and Ira Wasserman, Phys. Rev., {\bf D62}, 024011 (2000); C.
Csaki, M. Graesser, L. Randall and J. Terning, Phys. Rev., {\bf D62}, 045015
(2000); W. D. Goldberger, M. B. Wise, Phys. Lett. {\bf B475,} 275 (1999); C.
Charmousis, R. Gregory and V. Rubakov, Phys. Rev., {\bf D62}, 067505 (2000).

\bibitem{kamke}  E. Kamke, {\it Differential Cleichungen. Losungsmethoden
und Lonsungen: I. Gewohnliche Differentialgleichungen} (Leipzig, 1959).

\bibitem{kad}  B. B. Kadomtsev and V. I.\ Petviashvili, Dokl. Akad. Nauk
SSSR, {\bf 192,} 753 (1970).

\bibitem{dryuma}  V.\ S. Dryuma, Pis'ma JETP, {\bf 19,} 753 (1974).

\bibitem{zakhsh}  V. E. Zakharov and A. B.\ Shabat, Funk. Analiz i Ego
Prilojenia [in Russian] {\bf 8,} 43 (1974).

\bibitem{har}  B. Harrison, Phys.\ Rev. Lett., {\bf 41,} 1197 (1978).

\bibitem{lieb}  G.\ Liebbrandt, Phys. Rev. Lett. {\bf 41,} 435 (1978).

\bibitem{whith}  G. B. Whitham, J. Phys. A. Math., {\bf 12,} L1 (1979).

\bibitem{belinski}  V.\ A. Belinski and V.\ E. Zakharov, JETP. {\bf 75},
1953 (1978).

\bibitem{korn}  G. A. Korn and T. M. Korn, {\it Mathematical Handbook}
(McGraw--Hill Book Company, 1968).
\end{references}
\end{document}